\documentclass[a4paper]{jpconf}
\usepackage{graphicx}
\begin{document}
\title{Precision lattice QCD calculations and predictions of fundamental 
physics in heavy quark systems}

\author{Christine Davies}

\address{Department of Physics and Astronomy, University of Glasgow, Glasgow, G12 8QQ, UK}

\ead{c.davies@physics.gla.ac.uk}

\begin{abstract}
I describe the recent success in performing accurate calculations of the 
effects of the strong force on particles containing bottom and 
charm quarks. Since quarks are never seen in isolation, and so cannot be 
studied directly, numerical simulations are key to understanding the properties 
of these particles and extracting information about the 
quarks. The results have direct impact on the worldwide 
experimental programme that is aiming to determine the parameters of 
the Standard Model of particle physics precisely and thereby uncover 
or constrain the possibilities for physics beyond the Standard Model. 
The numerical simulation of the strong force is a huge computational 
task and the recent success is the result of international 
collaboration in developing techniques that are fast enough to 
do the calculations on 
powerful supercomputers. 

\end{abstract}

\section{Introduction - the physics problem}

The aim of particle physics is to uncover the fundamental particles and 
interactions that drive physics at the smallest distance scales. 

This is often represented as `peeling away the layers of an onion' starting 
with the atom as the outside layer. 
In fact it is rather more complicated. The subunits of the atomic nucleus have 
been known for seventy-five years. They are called protons and neutrons and 
are very similar particles although the proton has electrical 
charge equal to, but opposite in sign, to that of the electron 
and the neutron is electrically neutral. 
Protons and neutrons can be relatively easily isolated and studied in the laboratory 
and their mass and charge determined. Indeed protons accelerated to very high 
energies are one of the key experimental tools in studying particle physics at 
laboratories such as Fermilab and CERN. 
When high energy protons collide, a huge myriad of other particles is 
produced that can be tracked in particle detectors and their electrical charge 
and mass measured. These particles are not, however, the subunits of the proton. 
In fact, apart from the electron and its partners the muon and tau, they are  
particles made themselves of the same subunits as the proton. These subunits 
are known as quarks and particles made of them are collectively called hadrons. 
We have never been able to isolate quarks and so 
can only infer their existence from the results of these experiments using
theoretical understanding. The existence of quarks is now universally accepted, 
but the problem that they
are never seen as free particles makes it even harder to `peel away another 
layer of the onion'. A key requirement for progress on this will be precise determination 
of the properties of quarks and, because of the nature of quarks, this requires 
both precise experimental results {\it and} precise theoretical calculations. 
Lattice QCD aims to provide these theoretical calculations. 

\subsection{The Standard Model}

The Standard Model of particle physics describes the particles and interactions
at the most fundamental level currently known. There are three forces described by 
the Standard Model. The simplest and the one that we have most every day 
experience of is electromagnetism. At the every day level we can use the 
classical theory of Maxwell's equations that deals with electric and magnetic fields 
and the interaction of electrically charged particles with those fields. 
When we are dealing with the 
small distances of the subatomic world we must convert this to a 
quantum theory that can handle phenomena such as the production of particles 
and antiparticles out of the energy of an electromagnetic field. 
In the quantum theory electrically charged particles, such as the electron, 
are described by a quantum field, $\psi(x)$. The particles interact by 
exchange of the quanta of the electromagnetic field, called photons, with 
quantum field $A_{\mu}(x)$. The Lagrangian of the theory is particularly 
elegant, possessing a symmetry called a {\it local gauge symmetry}, that 
so limits the types of interactions that are allowed, and constrains the 
theory to be well-behaved, that it can be taken as a requirement rather 
than a result. The fact that the other two forces described by the Standard 
Model, the weak force and the strong force, are also {\it gauge theories} 
with a local gauge symmetry is then very satisfying. It took some 
time to realise that this was true, however, because at first 
sight the weak and strong forces look very different from electromagnetism. 

The strong force operates inside the atomic nucleus between the protons 
and neutrons and keeps the nucleus together against the electrical repulsion 
of the protons. The electrical force operates over relatively large 
distance scales  - it keeps the electrons attracted to the atomic 
nucleus in an atom after all. Playing with fridge magnets is a simple 
demonstration of this fact in the every day world. 
The long range is associated with the fact that the photon is a massless 
particle, a consequence of the gauge symmetry of the theory. Instead the strong force is 
limited to the short range of the atomic nucleus and therefore 
seems to require a massive exchange particle. The interaction between 
protons and neutrons is not fundamental, however, but is the 
`left-over' result of the interaction between the quarks inside the 
protons and neutrons.  

\begin{figure}[h]
\includegraphics[width=18pc]{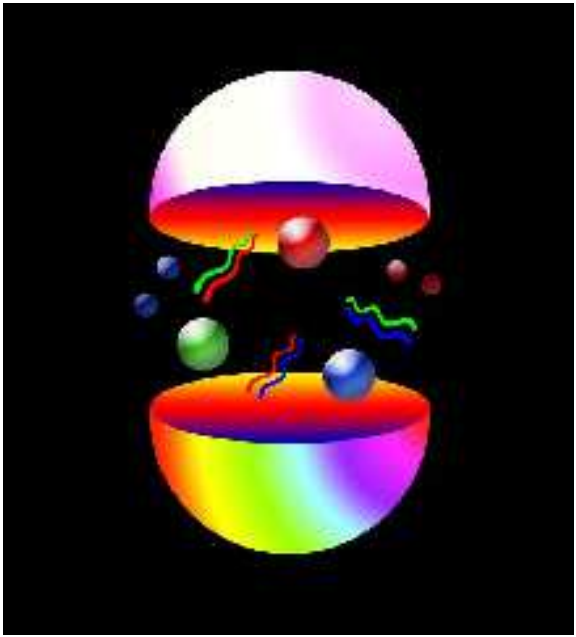}\hspace{2pc}%
\begin{minipage}[b]{18pc}\caption{\label{fig:proton}
The proton is made of 3 valence quarks (shown as 
large spheres) - 2 $u$
quarks and a $d$ quark in a complicated  and 
strongly interacting soup of gluons and sea quark-antiquark 
pairs (of flavors $u$, $d$ and $s$ and shown as small spheres).
The whole object has no color charge and so can 
exist as a free particle, unlike the quarks or gluons 
themselves. 

\vspace{1em}

There are six quark flavors, arranged into 
three `generations', given below as 2-d vectors. 
Each generation has 
a quark of electrical charge $-e/3$ (the top entry in the vector) and a quark 
of charge $+2e/3$ (the lower entry). 

\vspace{1em}
 
%\begin{center}
$ \quad \quad \quad \quad \quad \displaystyle{d \choose u} {s \choose c} {b \choose t} $
%\end{center}

\vspace{1em}

}
\end{minipage}
\end{figure}

\subsection{QCD as the theory of the strong force}

It turns out that the strong force can be 
described by a gauge theory called quantum chromodynamics (QCD) and 
this theory is very similar to quantum electrodynamics (QED) except that 
the equivalent of electrical charge, called color charge, comes in 
three possibilities instead of just being a number. The gauge symmetry 
of QCD is then described by a different {\it gauge group} and this gives 
rise to the very different behaviour of the strong force.  
Quarks carry an electrical charge which is either -2/3 or +1/3 times 
the charge on a electron, $-e$, (the basic unit of electrical 
charge).  Their color charge however is a vector of length the basic 
unit of color charge, $g$, but in a three-dimensional space with 
axes called red, green and blue. The antimatter partners of the quarks, 
antiquarks, have opposite electrical charge (just as the antielectron, 
or positron, has charge $+e$) and color charge anti-red, 
anti-blue or anti-green. The exchange particle, called the gluon, 
is massless like the photon but also carries a color charge, whereas 
the photon has no electrical charge. The gluon in fact has both a color 
and an anti-color charge in eight combinations. This means that gluons, 
as well as quarks, can emit gluons. It is this possibility that is 
at the root of the counter-intuitive behaviour of the strong force.  

We are used to thinking of electrical charge as being a measurable and 
fixed quantity - some number of units of $e$. It is determined in principle by 
taking a test charge at some distance from the charge in question and measuring 
the strength of the interaction between them. In fact, at the small distances
of the subatomic world, the charge measured in this way depends on the 
distance at which it is measured. Electrical charge appears to get stronger 
as you get closer to it. This is a result, in QED, of `vacuum polarisation' 
by which photons, produced by energy fluctuations in the vacuum, decay into 
an electron-positron pair that screen the charge in the same way that charge 
is screened inside a dielectric medium. As you get closer to the charge the 
screening is less effective and the charge seen increases.  
For color charge the opposite result occurs - gluons in the vacuum can 
produce screening quark-antiquark pairs but also `antiscreening' gluons.
The antiscreening wins out over the screening so that color charge 
gets smaller as you get closer to it and the strong interaction between 
color charges gets weaker.  This is known as `asymptotic 
freedom' and Gross, Politzer and Wilczek showed that QCD demonstrated 
this property in 1973 (and shared the 2004 Nobel Prize).  

QCD seems to have the right properties to explain the basic 
features of the strong force.  Experiments in which electrons are fired 
with high energy at protons (therefore probing short 
distances inside the proton) show that the constituents of the proton 
are basically behaving like free particles with rather weak strong 
force interactions. When high energy protons collide together the hadrons 
that spray out from the interaction do so in a way that bears the 
imprint of a basic strong force interaction in which a quark-antiquark 
pair were produced and flew apart, perhaps weakly emitting gluons. We
cannot see this basic interaction directly because as the quark 
and antiquark separate more quarks, antiquarks and gluons are produced 
from the interaction energy so that only hadrons, bound states of 
quarks, antiquarks and gluons that have no overall color charge,  
can be seen in particle detectors (see Figure~\ref{fig:proton}). Theoretical calculations in 
QCD can be done analytically for these high energy, short distance, 
examples where the strong force is relatively weak. Then the so-called 
`coupling constant', $\alpha_s=g^2/4\pi$, is relatively small and 
a powers series expansion in the number of interactions, or powers 
of $\alpha_s$, makes sense. This is the standard approach, known 
as perturbation theory, for QED 
where $\alpha_{em}=e^2/4\pi$ is naturally very small and it allows 
us to test QED to fantastic precision. Good tests of QCD are also 
possible for specific results from high energy collisions that 
survive the conversion from quarks into hadrons.  

QCD can tell us more than this, however. In principle all the properties 
of hadrons are predictable from QCD once the free parameters of the 
theory, the quark masses and $g$, are fixed. In practice this is a very 
difficult problem since at the typical distance scales inside a hadron, 
the strong interaction is very strong and $\alpha_s$ very large. Perturbation 
theory is no guide since it is no longer true that including more 
interactions in the calculation (and therefore higher powers of $\alpha_s$) 
gives a smaller number. Instead any number 
of interactions is equally important and the theory becomes non-linear 
and very complicated. It has after all to generate the phenomenon by 
which quarks are confined within hadrons and cannot escape to be free 
particles. 
Numerical simulation is the only way to solve this 
and, as we shall see in the next section, is an enormous computational task.  

An issue that complicated the understanding of results from high energy 
experiments was the sheer number of different hadrons that there are, 
each one with a distinct mass and having various electrical charges and spins.
Some hadrons live for long enough to be seen distinctly in particle detectors
(such as protons); others live for only a short time, and their existence 
must be reconstructed from the tracks of longer lived decay products. 
We now know that the reason for this `particle zoo' is that there are six 
different types or `flavors' of quarks. They are dubbed `up', `down', `strange', 
`charm', `bottom' and `top' quarks, in order of increasing mass, and abbreviated 
by their initials: $u$, $d$, $s$, $c$, $b$, and $t$. The $d$, $s$ and $b$ 
quarks have electrical charge $-e/3$ and the $u$, $c$ and $t$ quarks have electrical 
charge $2e/3$. In fact the heaviest quark, the top 
quark, does not live for long enough to make interesting bound states, so 
the hadrons that we can study have as basic constituents 
different combinations of the five 
lighter quarks. This gives a very rich physics because the quarks can be 
put together in many combinations and within those combinations different 
spin orientations and different amounts of orbital angular momentum give 
different hadrons. All of the hadrons should be described by QCD, however, 
and as we shall see it can be simpler to do QCD calculations for some 
of the more esoteric short-lived hadrons than for the long-lived every day 
proton. Here I will describe mainly results for particles made of charm 
and bottom quarks.   

\subsection{The weak force}

As well as the electromagnetic and strong forces quarks are also subject to 
the third force of the Standard Model, the weak force. This force is best 
known for its effects in nuclear beta decay. In that process the atomic nucleus 
of one element changes to that of another and  emits a beta ray (an electron). 
In fact what is happening is that a down quark in a neutron in the original 
nucleus changes to an up quark and emits a $W-$ boson. 
Since a neutron has as basic constituents 
two  $d$ quarks and a $u$, $ddu$, and the proton is $uud$ then the neutron changes 
to a proton in this process. The original nucleus thus changes to that of the element 
with one higher atomic number. The $W$ boson is the exchange boson of the weak 
force and that decays to an electron, which is detected, and an anti-neutrino, which
is very hard to see. The weak force was also originally hard to fit into the pattern 
of a gauge theory set by QED because the $W$ boson has a mass. However, again this 
was misleading and the result of spontaneous breaking of the weak force gauge 
symmetry caused by 
the Higgs boson.

Weak force interactions are the only way for a quark 
to change flavor. Since the $W$ boson is electrically charged with charge 
$+e$ for $W^+$ and $-e$ for $W^-$, when a quark emits a $W$ boson it 
changes flavor from one of the three $+2e/3$ quarks to one of the $-e/3$ set or vice 
versa. 
The existence of the Higgs boson and the quark interactions with that 
boson mean that the coupling between the different quark flavors and 
the W boson are not all the same but appear as a $3\times 3$ matrix (the 
Cabibbo-Kobayashi-Maskawa or CKM matrix) that can 
have complex numbers for coefficients. This has the interesting consequence 
that the symmetry between matter and antimatter can then be broken. 
We know that this symmetry is broken in the real world since we live 
in a universe made of matter which presumably started from a symmetric 
equal mixture of matter and antimatter at the Big Bang. However what we 
don't know is whether the description of this breaking through the
$3\times 3$ CKM matrix is the correct one. 
The first stage in this process is to determine the elements of this 
matrix accurately and in particular to discover whether it is a unitary 
matrix as required by the Standard Model picture. The matrix connects the 
$+2e/3$ quarks to the $-e/3$ charged quarks with entries:
\[ \left( \begin{array}{ccc}
{\bf V_{ud}}  & {\bf V_{us}} & {\bf V_{ub}} \\
{\bf V_{cd}} & {\bf V_{cs}} & {\bf V_{cb}} \\
{\bf V_{td}} & {\bf V_{ts}} & {\bf V_{tb}} \\
\end{array} \right) \]  \\
If this matrix turned 
out not to be unitary, it would be the `smoking gun' of physics beyond 
the Standard Model. 

A large part of the current particle physics experimental programme is 
devoted to this effort.  It turns out that critical elements of this 
matrix can be obtained by the study of the weak interactions of the 
$b$ quark. The $b$ quark will
emit a $W^-$ boson that can decay to an 
electron and antineutrino by analogy to the $d$ quark interaction
that is nuclear $\beta$ decay described above. 
At a simple level the rates of the weak decay of the 
$b$ quark to the different $2e/3$ quarks are proportional to the 
squares of the appropriate elements of the CKM matrix. 
However, as described earlier, $b$ quarks cannot be found 
in isolation but only bound into hadrons. The QCD interactions
inside the hadron affect the decay rate and must be calculated 
precisely in order to extract the appropriate CKM matrix element 
from the experimental result. This must be done in lattice QCD 
and, as we shall see, is one place where lattice QCD is now 
making a big contribution and will provide key results in the near 
future. 

\section{Introduction - the computational problem}

A quantum field theory can be defined by its Feynman path integral 
and this is a convenient way to express it for numerical simulation. 
This formulation has a lot of overlap with statistical mechanics and shares 
some common language. 
The basic integral is often called the partition function and for a 
gauge theory is expressed as:
\begin{equation}
Z = \int dA_{x\mu} d\psi_x d\overline{\psi}_x exp(iS(A,\psi,\overline{\psi}))
\end{equation}
where the integral is over the field variables of the gluon ($A_{\mu}$) and quark fields 
($\psi, \overline{\psi}$)
at every point in 4-d space-time, $x$. $S$ is the action of the theory, the 
space-time integral over the Lagrangian and a function of the gluon and 
quark fields. The advantage of expressing the theory in this way is that 
the left-hand side is a quantum transition amplitude, in fact from the vacuum state 
to the vacuum state, but the right-hand side expresses this as an integral over 
classical field variables that can be represented as numbers on a computer. 
The integral sums over all paths (configurations of the $A$ and $\psi$ fields) 
between the initial and final quantum 
states weighting each path with $e^{iS}$. 
In the continuous space-time of the real world this integral 
is not well-defined because it has infinitely many dimensions if there are 
infinitely many space-time points and the field variables are unconstrained. 

\begin{figure}[h]
\includegraphics[width=16pc]{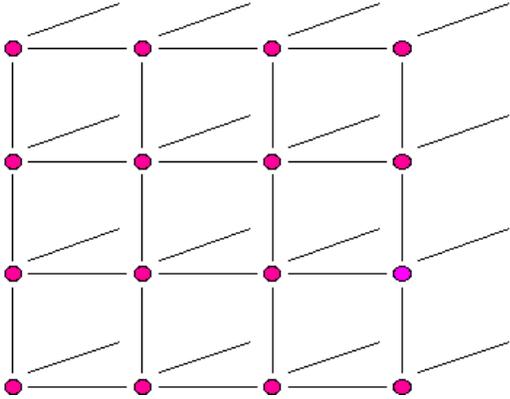}\hspace{2pc}%
\begin{minipage}[b]{18pc}\caption{\label{fig:lattice}
A 2-dimensional rendition of a 3-dimensional lattice. Lattice QCD uses 4-dimensional 
lattices, usually periodically joined into a 4-dimensional torus. 
The spacing between the points (and therefore the smallest 
non-zero distance on the lattice) is denoted $a$. The largest momentum 
that can be represented meaningfully on the lattice is $\pi/a$ so 
the lattice provides an ultraviolet cut-off on momenta.

}
\end{minipage}
\end{figure}

To adapt the integral for numerical solution several steps are needed: 
\begin{itemize}
\item{take a finite box of space-time, of length $L$ on a side}
\item{split the box up into a finite number of points in a 4-dimensional lattice grid. 
Call the spacing of the grid the lattice spacing, $a$, see Figure~\ref{fig:lattice}. The gluon field variables live on the 
links of the lattice and the quarks on the sites. }
\item{Discretise the interactions of the QCD Lagrangian onto the space-time lattice. The integral is now finite. }
\item{rotate time to imaginary time so that the oscillatory factor $e^{iS}$ becomes 
the exponential factor $e^{-S}$. Now paths with large action are exponentially 
suppressed and we can compute the integral using the methods of importance sampling. }
\end{itemize}  

The simplest quantum transition amplitudes that we are interested in calculating 
involve making hadrons out of the vacuum, propagating those hadrons for a 
certain length of lattice time that allows their energy to be determined, and then 
destroying them back into the vacuum. This amplitude is then given by
\begin{equation}
\langle 0 | H(T) H^{\dag}(0) | 0 \rangle  = \frac{1}{Z} \int dA_{x\mu} d\psi_x d\overline{\psi}_x H(\psi, \overline{\psi})_T H^{\dag}(\psi, \overline{\psi})_0 \exp(-S(A,\psi,\overline{\psi})).
\end{equation}

A further important adaption must now be made because of the nature of quarks. 
They are fermions (particles with half-integer spin) and therefore obey 
fermion statistics. This means that their field variables must be anticommuting 
numbers that cannot be represented on a computer (at least not without 
enormous difficulty). Instead we must integrate over the $\psi$ and  $\overline{\psi}$ 
fields by hand and this can be done because of the simple way that these 
fields appear in the Lagrangian. They appear in the form $\overline{\psi} M \psi$ 
where $\psi$ is a vector representing the quark field on every site of the lattice 
and $M$ is an (enormous) matrix linking every site on the lattice to every 
other site, and a function of the gluon fields. 
Then the quantum transition amplitude above becomes (for a particularly simple 
hadron $H$ which is a meson made of a quark and an antiquark):
\begin{equation}
\langle 0 | H(T) H^{\dag}(0) | 0 \rangle  = \frac{1}{Z} \int dA_{x\mu} {\rm Tr} (M^{-1}_{0T} M^{-1*}_{0T}) \det{M} \exp(-S_g(A)).
\label{eq:calc}
\end{equation}
The integral is now over the gluon fields only. 
$S_g$ is that part of the QCD Lagrangian that 
depends only on the gluon fields and $M$, as explained above is the quark interaction 
matrix that is also a function of the gluon fields. $M^{-1}$ is its inverse, 
$\det{M}$ its determinant 
and Tr is a trace over various matrix indices. 

How we proceed to calculate this ratio of path integrals is by first generating sets 
of `configurations' of gluon fields with a probability related to how 
much they will contribute to this integral. We can think of them 
as `typical snaphots of the QCD vacuum'. A gluon field configuration is 
a complete set of values for the gluon field on every link of lattice. The 
probability distribution for the configurations is $\det{M}\exp(-S_g)$.
Starting with a random configuration, and using algorithms such as the 
hybrid Monte Carlo algorithm, we generate a long Markov chain of configurations
which converges to this probability distribution.
We can then record a configuration every so often to make an ensemble 
of configurations with the right distribution. `Every so often' means 
a spacing in Markov chain time that allows the configurations to be 
reasonably independent of each other. Typically there would be 
five hundred or so configurations in the ensemble. 

We then move to the `data analysis' stage of the calculation.
Many people can calculate different quantities of the form given in 
equation~\ref{eq:calc} from the same ensembles of gluon field 
configurations so the process divides naturally into two rather like, 
for example, a particle physics experiment in which the generation 
of the data and its analysis are separate but linked activities. 
At the data analysis stage, the integral in equation~\ref{eq:calc} 
is evaluated by simply `measuring' (ie. calculating) the value of 
${\rm Tr} M^{-1}_{0T}M^{-1*}_{0T}$ on every gluon field configuration 
and averaging over the ensemble. The factors of $1/Z \det{M} e^{-S_g}$ 
are taken care of by the probability distribution of the configurations 
in the ensemble, provided we have a good statistical sample. 
There will be statistical errors associated with the number of 
independent configurations we have generated. Systematic errors in the result 
will be discussed below. 

An integral like equation~\ref{eq:calc} has an expected form as
a function of $T$ that we can then fit to extract quantities 
like the mass of hadron $H$, and amplitudes for simple decay 
processes (see the next section for what kind of decay processes 
these are). It is important to realise that the raw results 
are all in units of the lattice spacing, `lattice units'. The lattice spacing does 
not appear explicitly anywhere in the calculation but is 
implicitly controlled by the parameter $g$ in the QCD 
Lagrangian. We first have 
to determine what it is from the results of one calculation before 
we can convert the results for other calculations from lattice 
units to physical units. 
Physical units for masses would be 
kg but as particle physicists it is natural to use energy equivalents
in electron-volts, where the energy equivalent of the mass 
of a proton is then roughly 1 ${\rm GeV}$. 
From fitting a hadron correlator (the result for equation~\ref{eq:calc})
we obtain a hadron mass in the form $m_Ha$ in lattice units. Given a 
number for $m_Ha$ and an energy value for $m_H$ from experiment in ${\rm GeV}$ 
we can obtain $a^{-1}$ in ${\rm GeV}$. Then, using this 
value of $a^{-1}$ results for other hadron masses obtained from 
the correlators in the form of equation~\ref{eq:calc} for other hadrons 
can be converted from lattice units to 
physical units and compared with experiment. 

\subsection{Discretisation errors}

The use of a lattice and a lattice spacing is simply a device
to make the calculation tractable and results in physical units 
should not depend on the lattice spacing to be reliable. 
Of course the results do depend on $a$ because of {\it 
discretisation errors}. These errors arise when the QCD 
Lagrangian is discretised on to the lattice in the same way 
that they do when all differential equations from 
continuous space-time are discretised onto a lattice for 
numerical solution.  For example, a derivative can be 
discretised as a simple finite difference. This is 
correct as $a \rightarrow 0$ but at finite values of $a$ we 
can expand the difference in terms of continuum derivatives 
to see that there are errors at $\cal{O}$$(a^2)$. 
\begin{equation}
\frac{\partial \psi(x_j)}{\partial x} = \frac{\psi(x_j + a) - \psi(x_j -a)}{2a} + {\cal{O}}(a^2)
\end{equation}

Most calculations are done at values of 
$a$ around 0.1fm which corresponds to $a^{-1}$ of 2 GeV. 
A typical scale for discretisation errors might be 
0.5 GeV in which case an $a^2$ error might be expected 
to be $(0.5/2)^2$ = 6\%. This error is rather too large
to accept for a precision calculation, even allowing 
for the possibility of obtaining results at several 
values of $a$, fitting the $a$ dependence and so 
removing it with some error.
One problem is that going to smaller values of $a$ 
rapidly becomes extremely expensive as the cost of 
the calculation grows as something like $a^{-7}$. 
A lot of work was done during the 1990s to study 
discretisation errors in lattice QCD and reduce them further. 

In a quantum field theory like QCD further improvement 
of discretisation errors is not completely straightforward. 
The usual solution would be to remove the $a^2$ error 
by adding terms from a higher order differencing scheme 
to cancel it. This removes the leading $a^2$ error 
in lattice QCD but there are sizeable subleading errors.  
The point of an improved discretisation scheme is to 
make the equations more closely resemble those in continuous 
space-time. For QCD, however, the quark fields are radiating 
gluon fields at all length scales. The difference then between 
the lattice discretisation and the continuum equations is 
affected by the fact that in continuous space-time very 
short-distance interactions can occur, with length scale 
smaller than the lattice spacing, that are not allowed 
in lattice QCD. Luckily we know that in this regime the 
QCD coupling constant becomes rather weak and so we 
can calculate this difference, and adjust for it, as 
a power series expansion in $\alpha_s$. This has allowed 
a very successful programme of the improvement of 
discretisation errors in lattice QCD. The results you 
will see in the results section have applied this programme 
to reduce discretisation errors to the few percent level, 
and to quantify the remaining errors by looking at 
results at more than one value of $a$. 

\subsection{Valence quarks and sea quarks}

The reason that lattice QCD calculations are 
so computationally demanding is the presence of the 
factors of $M^{-1}$ and $\det{M}$ in equation~\ref{eq:calc}. 
These factors come from the quark fields in QCD but 
have effectively two somewhat different sources. 
We say that the $M^{-1}$ factors come from the 
`valence quarks' and the $\det{M}$ factors come from 
the `sea quarks'. The picture in Figure~\ref{fig:proton} should 
make this distinction clear. The basic constituents 
of a hadron are 3 quarks if it is a baryon and a quark 
and an antiquark if it is a meson. We have already 
discussed the neutron as being made of $ddu$ and 
the proton as being $uud$ (both are baryons). These quarks 
are known as the valence quarks and they dominate the 
properties and behaviour, for example in terms of 
decay modes, of the hadron. However, these quarks are 
living in a strongly interacting environment inside the 
hadron. This environment will contain lots of gluon 
radiation and quark-antiquark pairs made from this 
gluon radiation. These quark-antiquark pairs are known 
as the sea quarks. We have already discussed how these
quark-antiquark pairs screen color charge and how that is 
offset by gluon radiation that has an antiscreening effect.
If the sea quarks were missing, it is clear that there would 
be too much antiscreening and we could expect that 
the coupling constant $\alpha_s$ would not vary in 
the correct way with distance and this would 
introduce errors. 

The inclusion of the $\det{M}$ factor is enormously 
costly in terms of computer time. Early lattice 
calculations missed it out in an approximation known 
as the `quenched approximation'. This was indeed wrong, 
as described above, but it took some while to pin this 
down because of the lack of understanding of discretisation 
errors in the early days. Now it is clear that the errors
from using the quenched approximation are of order 20\% 
(see Figure~\ref{fig:ratio} in the section on results). 

The new lattice calculations include sea quarks 
and are a huge improvement in terms of accuracy.
The important sea quarks are those of lightest 
mass, $u$, $d$ and $s$, since they can most easily be 
made from energy fluctuations in the vacuum. Unfortunately 
the cost of including $\det{M}$ grows as the quark 
mass gets smaller so for the particularly light 
$u$ and $d$ quarks we have to operate in the calculation 
with masses above their physical values and extrapolate 
down to the real world result, guided by theoretical 
expectations. If we are close 
enough to the physical point, this extrapolation
does not introduce significant error. 

The computing cost for including $\det{M}$ is very large. 
The improved staggered formulation is the most 
efficient formulation we have for doing this and also 
the one with the most mature results (to be 
described in the next section). For that formulation 
ongoing calculations to generate an ensemble of 
gluon field configurations including sea $u$, $d$ and 
$s$ quarks on a $48^3\times 144$ lattice with spacing 
$a = 0.06$fm will take 0.5 - 1.5 Tflopyears depending 
on the $u/d$ quark mass.

\section{Results from lattice QCD}

The MILC collaboration has made ensembles of gluon configurations 
including 2+1 flavours of sea quarks. 
The `2' refers to sea $u$ and $d$ quarks which 
are taken to have the same mass for simplicity; the `1' refers to the 
sea $s$ quark. The formalism used for the quarks is the improved 
staggered one~\cite{asqact}, also known as asqtad. 
The sea $u/d$ quarks have various masses (for 
different ensembles) down to $m_s/10$ which 
is about a factor of two from the real world, and much lower than previous 
calculations. It is then possible to extrapolate to the real world 
value of the $u/d$ quark mass guided by chiral perturbation theory which 
dictates the behaviour of quantities as the $u/d$ quark mass goes to zero, 
and should be valid for sufficiently small $u/d$ mass. There is no numerical 
problem in reaching the correct $s$ quark mass but, since it is only possible 
after the calculation to work what the $s$ quark mass was, ensembles with 
different values of the $s$ quark mass are needed for interpolation 
purposes. 

\begin{figure}[h]
\includegraphics[width=18pc]{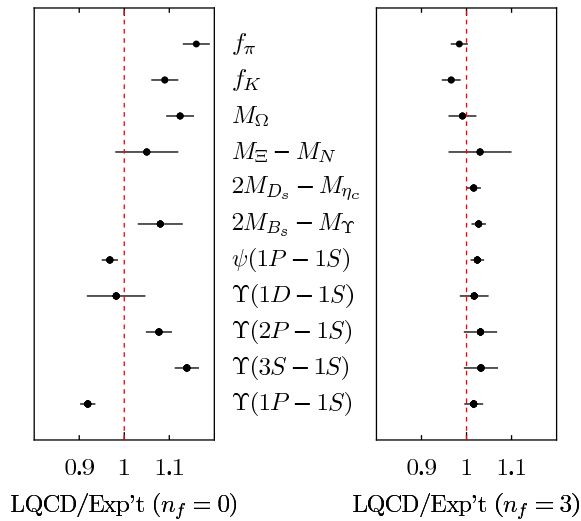}\hspace{2pc}%
\begin{minipage}[b]{18pc}\caption{\label{fig:ratio}
Lattice QCD divided by experiment for eleven different quantities 
for a range of hadrons. 
From the top: the decay constants of the $\pi$ ($u\overline{d}$) 
and $K$ ($u\overline{s}$) mesons; 
the $\Omega$ ($sss$) baryon mass; 
a mass difference between the $ssu$ baryon and 
the proton ($N$) which is insensitive to the $u/d$ mass; 
the difference between twice the $D_s$ ($c\overline{s}$) meson mass 
and the $\psi$ mass; the difference between twice the 
$B_s$ ($b\overline{s}$) meson mass and the $\Upsilon$ ($b\overline{b}$) mass; 
the orbital excitation energy 
in the $\psi$ ($c\overline{c}$) system; 
various excitations energies in the $\Upsilon$ system.
On the left, results in the quenched approximation. On the right, results 
on the MILC configurations
including sea quarks. }
\end{minipage}
\end{figure}

There are three sets of ensembles with three different values of the 
lattice spacing: the `fine' set with $a \approx$ 0.09 fm; the `coarse' 
set with $a \approx$ 0.12fm and the `supercoarse' set with $a \approx$ 0.18fm. 
This enables a test of the remaining discretisation errors. 
At each value of the lattice spacing, as $m_{u/d}$ is changed, the other 
parameters in the action are adjusted to keep $a$ roughly the same. 
This is important to avoid confusing effects, including discretisation 
errors, from changing $a$ with 
effects from changing $m_{u/d}$. 
The super-coarse configurations have $16^3\times 48$ lattice points, 
the coarse $20^3\times 64$ and the fine $28^3\times 96$ in general. 
The physical volume in all cases then is around 
$(2.5 {\rm fm})^3$ in the spatial directions 
which should be large enough to avoid significant errors 
from not having a big enough box of space-time. 
Larger physical volumes have been 
used for the lightest $u/d$ masses where these effects become more important. 

The analysis on these configurations has been done by the Fermilab, HPQCD, 
MILC and UKQCD collaborations. Most of the results come from the 
`fine' and `coarse' ensembles.  
QCD has five parameters for this situation: four quark masses and a 
coupling constant. These parameters can be fixed
by direct comparison of five hadron masses to experiment. 
The hadron masses used should 
be sensitive to the parameter being fixed but preferably not to any 
of the others. First the lattice spacing is fixed from the difference 
in mass between two mesons with the valence configuration $b\overline{b}$. 
These mesons are the `ground-state' of this system, called 
the $\Upsilon$ and its first radial excitation, the $\Upsilon^{\prime}$. 
This mass difference is useful because it is in fact very insensitive to (all) quark masses.
Then the $u/d$ quark mass is fixed from $M_{\pi}$, the $s$ quark 
mass from $M_K$, the $c$ quark mass from $M_{D_s}$ and the $b$ 
quark mass from $M_{\Upsilon}$. Further hadron 
masses and simple properties like decay constants 
can then be calculated as a precision test both of 
lattice QCD and of QCD itself in this strongly coupled regime. 
The results are shown for eleven different quantities covering 
the enormous range of QCD physics
in Figure~\ref{fig:ratio} (an updated version of that in reference~\cite{ratio}). 
Note that there are no free parameters 
in this plot since they have been fixed, as described above, by five quantities 
that are not shown.  

Each plotted point in Figure~\ref{fig:ratio}
is the ratio of the lattice QCD result to experiment, and the line 
marks the correct result i.e. 1. On the left are results 
using the quenched approximation in which sea quarks are ignored. It 
is clear then that a number of the results have significant (10-20\%) errors. 
On the right are results from the MILC 
configurations. Now the results agree with experiment across the board
with small errors. This also means that the QCD parameters are 
unambiguous, as they must be for real QCD  - 
any of the quantities on the plot could have been used 
instead to determine the parameters and the same result would have been 
obtained. 

\subsection{Determining the parameters of QCD}

Lattice calculations provide a very direct and accurate way to determine the parameters
of QCD: the coupling constant and the quark masses. These parameters 
come from some `theory of everything' and so 
can only be determined currently from comparison of QCD to experiment. This 
means giving up a number of predictions for hadron masses, as described above,
equal to the number of parameters. The power of QCD is the huge range 
of physics addressed with so few parameters, and knowing the parameters 
accurately can constrain theories of their origin. 

\begin{figure}[h]
\begin{minipage}{18pc}
\includegraphics[width=16pc]{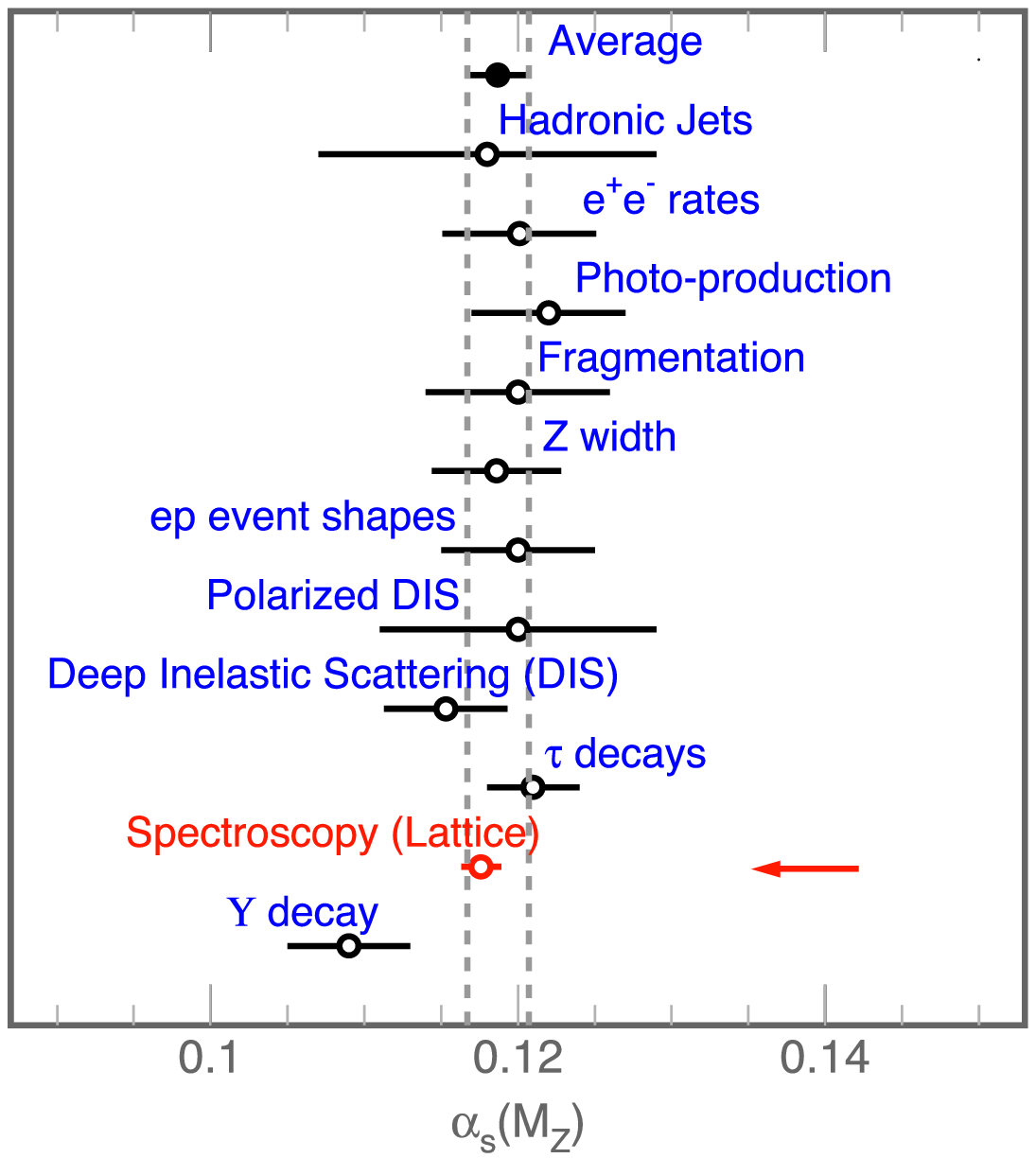}
\caption{\label{fig:alpha}The determination of the QCD coupling constant, $\alpha_s$, 
at a reference scale $M_Z$ from
a number of methods. The lattice QCD result of 0.1170(12) is the most accurate.}
\end{minipage}\hspace{2pc}%
\begin{minipage}{18pc}
\includegraphics[width=18pc]{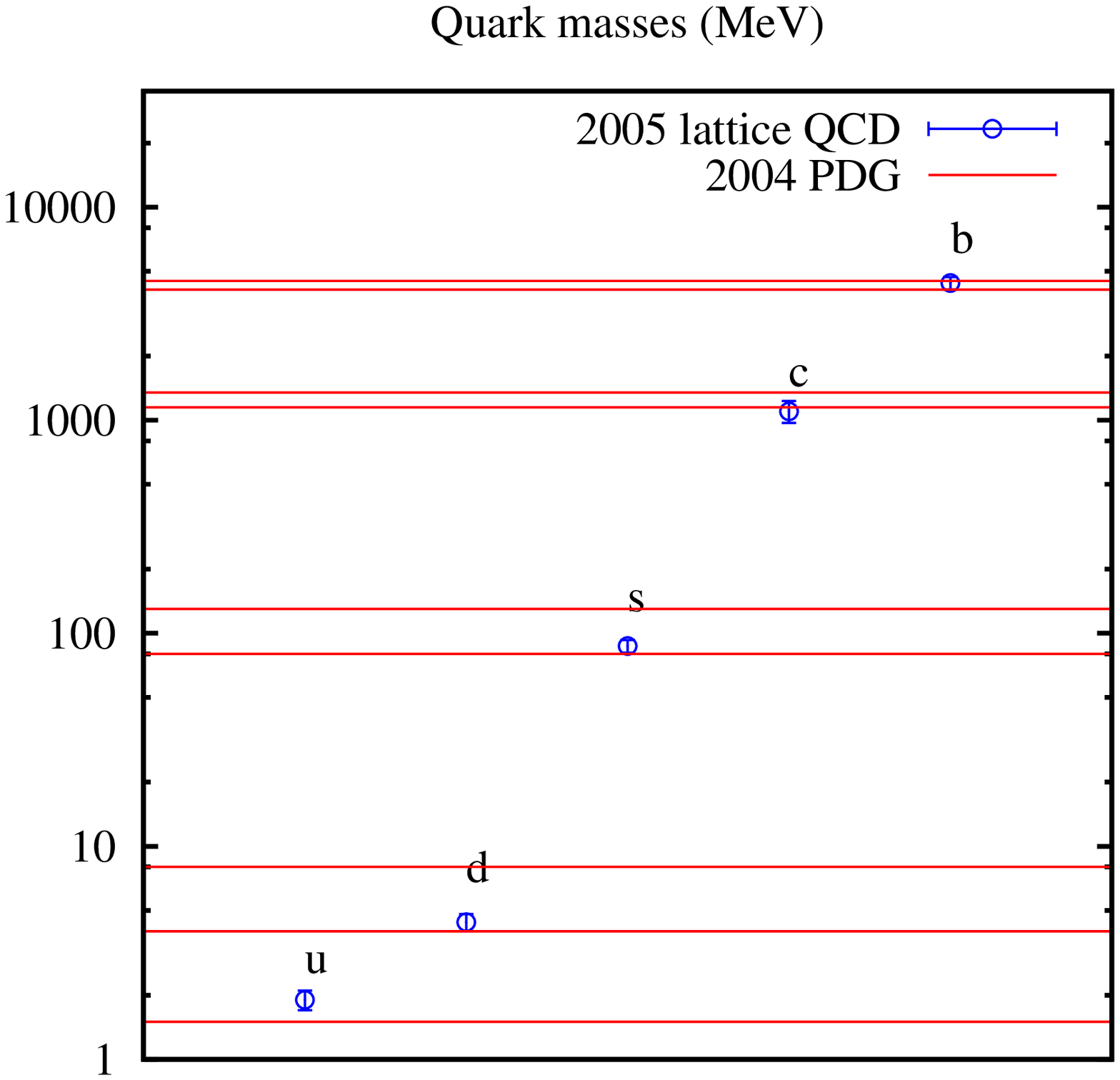}
\caption{\label{fig:qmass}Quark masses from lattice QCD compared to results obtained 
from other methods quoted in the 2004 Particle Data Tables (red lines). For reference 
the proton mass is 1000 MeV. }
\end{minipage} 
\end{figure}

The value of the QCD coupling constant, $\alpha_s$, could be obtained 
simply from the determination of the lattice spacing $a$ and knowing 
the value of $g$ in the action. In fact it can be done more 
accurately by `measuring' little loops of gluon fields on the 
lattice (typically one or two lattice spacings wide) 
given a calculation of these loops as a power series 
in $\alpha_s$. This perturbative series is valid for these short-distance 
quantities for the small lattice spacing values of the fine 
and coarse MILC lattices but we do need a series out to third 
order in $\alpha_s$ which is a very hard calculation. 
The result is shown in Figure~\ref{fig:alpha}, with $\alpha_s$ 
converted to a value at a reference distance scale and compared to 
results using other methods of combining theory and experiment. 
The lattice QCD result is the most accurate to date~\cite{quentin}. 

The determination of quark masses from lattice QCD calculations is even 
more straightforward. Indeed lattice QCD calculations provide a very 
clear definition of the quark mass, which is otherwise rather hard to pin down 
when confinement does not allow free quarks to appear. 
We simply have to adjust the quark mass parameter 
in the lattice QCD action until the mass of a hadron, preferably one that depends 
strongly on the quark mass, agrees with experiment. 
In practice this means doing calculations 
at several values of the parameter and interpolating or extrapolating 
to the right value. 
Of course this 
requires that there has already been a determination of $a$ since 
hadron masses in lattice calculations are given in units of $a$ 
and so is 
the quark mass parameter in the action. 
There also needs to be a renormalisation of the quark 
mass to give the result for continuous space-time rather 
than the lattice. This requires the calculation of a
perturbative power series in $\alpha_s$ that takes account of 
gluons of high momentum in continuous space-time missing in lattice QCD. 
These calculations are hard and so far have been done 
to second order only for light quarks~\cite{qmasses}. The accuracy for 
heavy quarks is limited by the fact that only first order 
renormalisation calculations exist~\cite{upsgray, nobes} and this is something that 
will be put right in the near future since it is clear 
from Figure~\ref{fig:qmass} that lattice QCD calculations 
of quark masses are much more accurate than other methods.  

\subsection{Hadron masses} 

An excellent test of QCD, as shown in Figure~\ref{fig:ratio}, is 
to calculate the masses of hadrons that are well known experimentally. 
The system of mesons based on $b\overline{b}$, known collectively 
as the $\Upsilon$ system or bottomonium, are particularly good 
because there are many radial and orbital excitations that 
are relatively stable mesons and that can therefore be experimentally 
characterised very accurately. In lattice QCD we can also do very 
accurate calculations for this system with much less 
computer power than is necessary for the light quarks. Because the 
$b$ quarks are heavy, they move slowly inside their bound states 
and the part of the Lagrangian of QCD that describes their interactions 
can be simplified to take account of this in a method known as 
Nonrelativistic QCD. Figure~\ref{fig:ratio} shows the excellent 
results that can be obtained and this gives us great confidence 
that we can handle $b$ quarks well in lattice QCD~\cite{upsgray} (this 
will be important for the next subsection). Similar methods 
can be applied to the somewhat lighter $c$ quarks. $c\overline{c}$ bound 
states form the $\psi$ system or charmonium, for which the orbital 
excitation energy is compared to experiment in Figure~\ref{fig:ratio}. 
 
\begin{figure}[h]
\begin{minipage}{18pc}
\includegraphics[width=18pc]{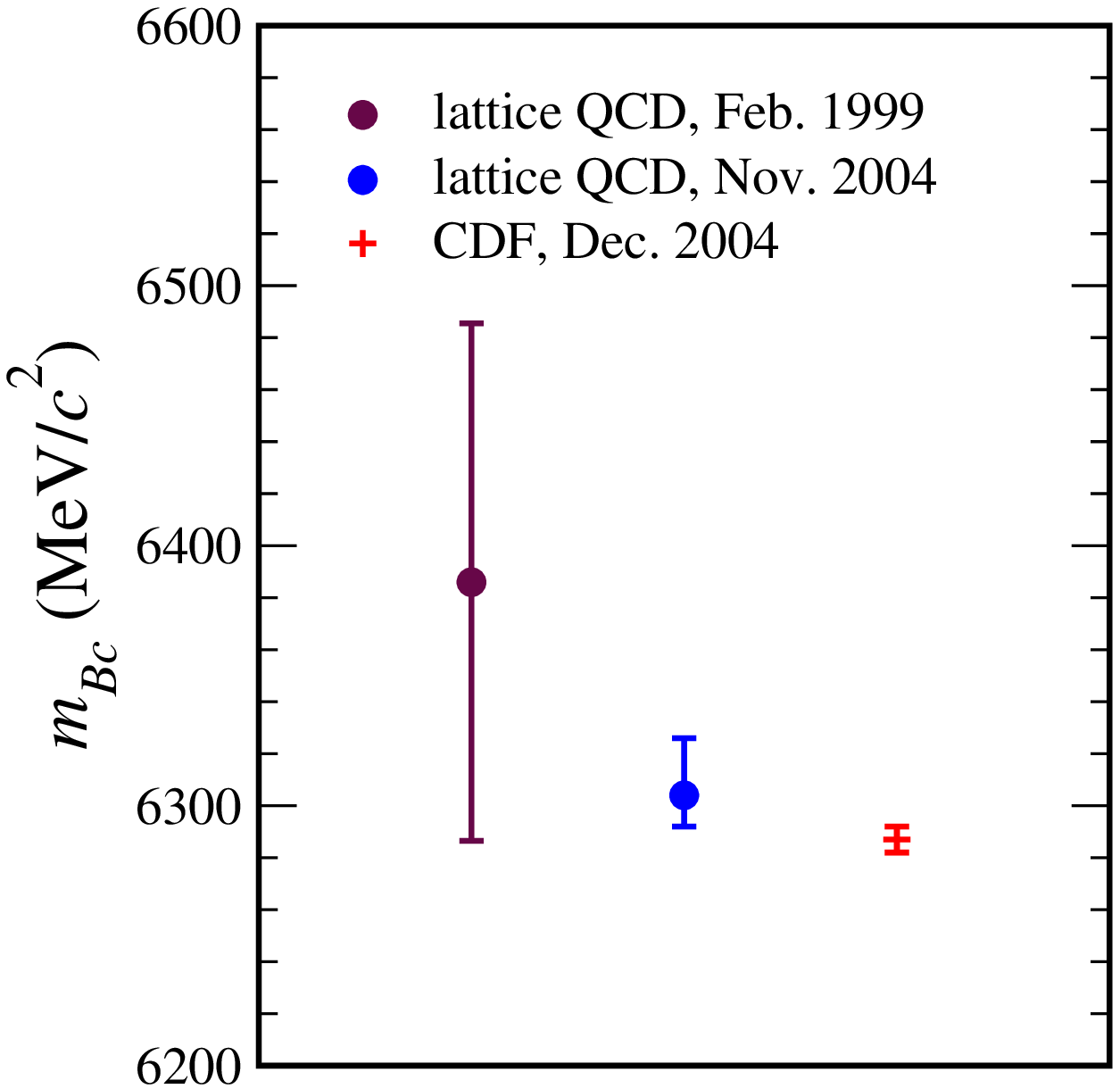}
\caption{\label{fig:bc}The lattice prediction for the mass of the $B_c$ meson has improved 
enormously from the old quenched calculation of 1999 to the new calculation 
that includes the effects of sea quarks. On the right is the experimental result that 
followed. }
\end{minipage}\hspace{2pc}%
\begin{minipage}{18pc}
\centerline{\includegraphics[width=10pc]{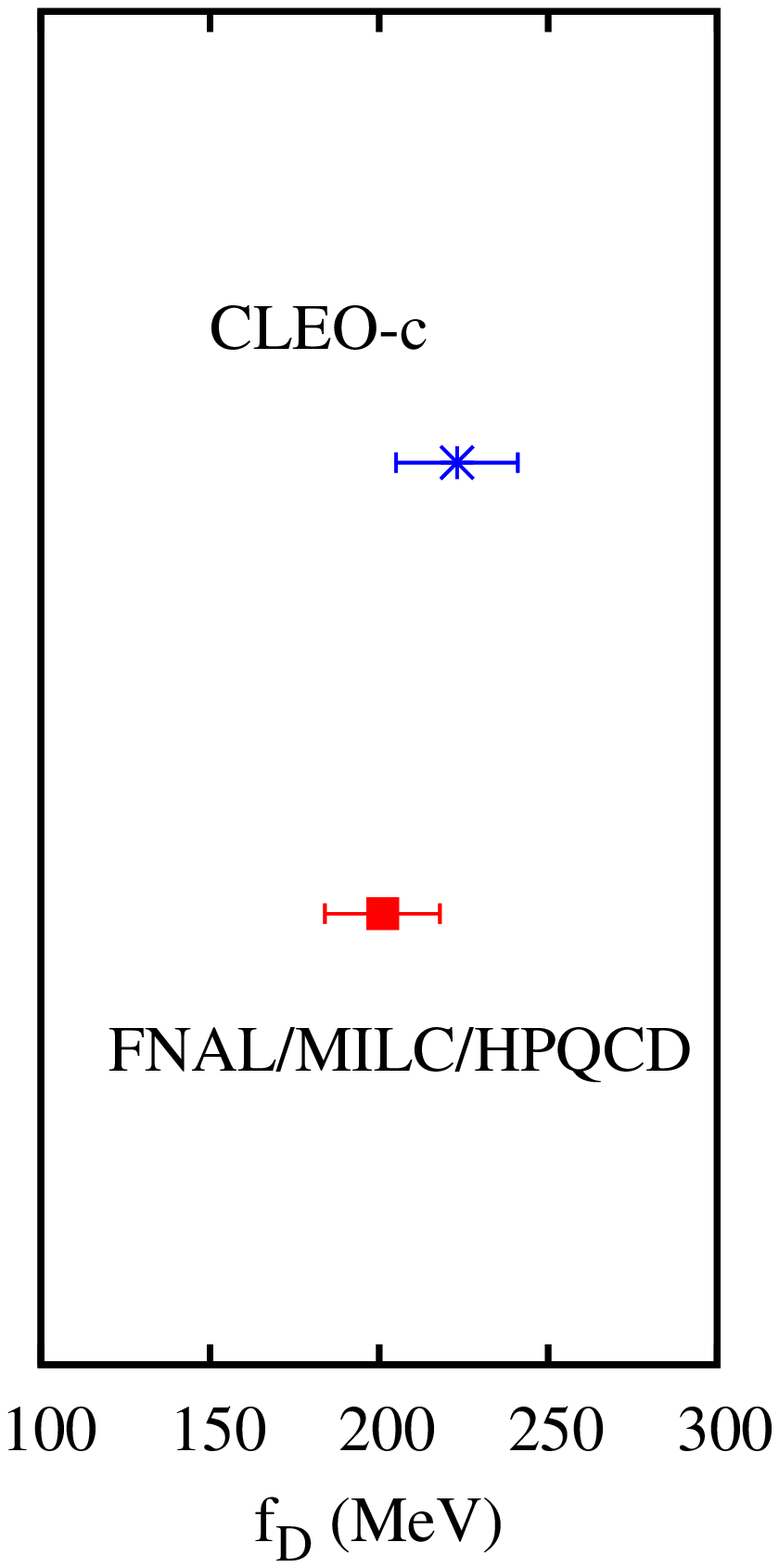}}
\caption{\label{fig:fD}Lattice QCD result for the decay constant of the $D$ meson~\cite{fD}
compared to the result obtained by dividing the experimental value of the 
leptonic decay rate of the $D$ from CLEO-c by a previous result for $(V_{cd})^2$.}
\end{minipage} 
\end{figure}

The success of these lattice calculations for the spectrum of 
$\Upsilon$ and $\Psi$ states enabled a lattice prediction for the 
mass of the $B_c$ meson to be obtained~\cite{allison} ahead of the experimental 
results announced by the CDF collaboration at the end of 2004~\cite{cdfbc}.  
The lattice prediction was made by calculating the difference 
in mass between the $B_c$ meson and the average of the 
$\Upsilon$ and $\Psi$ mesons. Taking this difference allows a 
number of systematic errors to cancel. The result obtained, 
6.304(20) GeV agrees well with the subsequent CDF 
result of 6.287(5) GeV and 
is a huge improvement over 
the accuracy that was possible in previous quenched calculations, 
see Figure~\ref{fig:bc}. Ongoing theoretical and experimental~\cite{cdfbc} 
work will improve these numbers further.

\subsection{Weak decay rates for hadrons}

Decay rates that can be accurately calculated for 
hadrons are those in which there is at most one 
hadron in the final state. Luckily a lot of the interesting 
decay modes caused by the weak force fall into this 
category. There are decay modes in which a quark and 
antiquark in a meson annihilate to a $W$ boson (obviously 
this requires the meson to have an electrical charge) that 
then decays to an electron and an antineutrino. This is 
known as a leptonic decay because the final products are 
all so-called leptons (particles that have no color charge and 
do not feel the strong force). There are also decay modes 
in which a quark or antiquark inside a hadron changes flavor 
and emits a $W$ boson. The meson therefore changes to another 
type and leptons are produced from the $W$. This is the analogue of 
nuclear $\beta$ decay and is known 
as semileptonic decay since there is one hadron, as well 
as leptons, in the final state. 

As described in subsection 1.3 an element of the CKM 
matrix appears every time a $W$ boson is emitted by a 
quark. The rate for leptonic or semileptonic decays 
is then a product of the square of the appropriate 
CKM element, say $V_{ab}$, multiplied by the rate of the 
basic process in which the quark-antiquark annihilation 
occurs (for leptonic decay) or the quarks and antiquarks 
rearrange themselves (for semileptonic decay). This basic 
process takes place in a background of strongly 
interacting QCD radiation and so needs the techniques of 
lattice QCD for its calculation. 
Then to determine $V_{ab}$ we simply divide the experimentally 
determined actual rate by the lattice QCD calculation of 
the basic process, and take the square root. 
Luckily there is an example of  decay mode of this kind 
for all the CKM elements in the first two rows of the matrix: 
\[ \left( \begin{array}{ccc}
{\bf V_{ud}}  & {\bf V_{us}} & {\bf V_{ub}} \\
\pi \rightarrow l\nu & K \rightarrow l \nu & B \rightarrow \pi l \nu \\
 & K \rightarrow \pi l \nu &  \\
{\bf V_{cd}} & {\bf V_{cs}} & {\bf V_{cb}} \\
D \rightarrow l\nu & D_s \rightarrow l \nu & B \rightarrow D l \nu \\
D \rightarrow \pi l\nu & D \rightarrow K l \nu &  \\
{\bf V_{td}} & {\bf V_{ts}} & {\bf V_{tb}} \\
\langle  B_d | \overline{B}_d \rangle  & \langle B_s | \overline{B}_s \rangle  & \\
\end{array} \right) \]
 As described earlier, the determination of the CKM elements and tests of 
the self-consistency of the CKM matrix are a big part of the current experimental 
programme and lattice calculations of these leptonic and semileptonic 
processes will be a key factor in 
the precision with which this can be done. For the decay rates that will 
be used for CKM element determination it is important to have cross-checks 
against experiment of other similar decay rates as a confirmation of the 
systematic error analysis. 

The CKM elements for $b$ quarks, $V_{ub}$ and $V_{cb}$, are the most 
poorly known. Lattice calculations are currently being done for 
the relevant leptonic and semileptonic decays for the mesons called 
$B$ mesons that contain a $b$ quark or antiquark and a $u$, $d$ or $s$ 
antiquark or quark respectively. These mesons are being studied 
extensively in the `$B$-factory' experiments at SLAC and KEK in Japan 
and will be studied further in the LHCb experiment at CERN. A good test of these 
calculations is the equivalent calculation for $c$ quarks, since 
a lot of the understanding of systematic errors in the lattice 
calculations is similar. For $c$ quarks the CKM elements are 
well-known from other processes. This means that an experimental result for 
the leptonic decay rate of the equivalent of the $B$ meson called 
the $D$ meson can be used with a known value of $V_{cx}$ 
to determine the annihilation amplitude or `decay constant', $f_D$, that can be 
directly calculated in lattice QCD. The CLEO-c collaboration have 
taken up the challenge of determining these leptonic rates for 
direct tests of the lattice calculations. Currently the situation is as in 
Figure~\ref{fig:fD} in which the lattice result~\cite{fD} and the experimental 
result~\cite{cleo} have roughly 8\% errors and agree within this precision.  
The experimental results will improve as the experiment runs 
for longer. The challenge for the lattice QCD theorists is to 
improve their discretisation of the QCD Lagrangian for charm 
quarks so that they can improve their prediction of $f_D$
to the few percent level
ahead of precise experimental results. In fact this year the 
Belle collaboration, working at KEK in Japan, produced a first result for the leptonic 
decay rate of the $B$ meson~\cite{belle}. In principle this can be used, along 
with a lattice calculation of $f_B$, to determine $V_{ub}$ but at 
present the experimental result is not accurate enough for this. 
It shows however, the possibilities that might exist in the future. 

\begin{figure}[h]
\includegraphics[width=30pc]{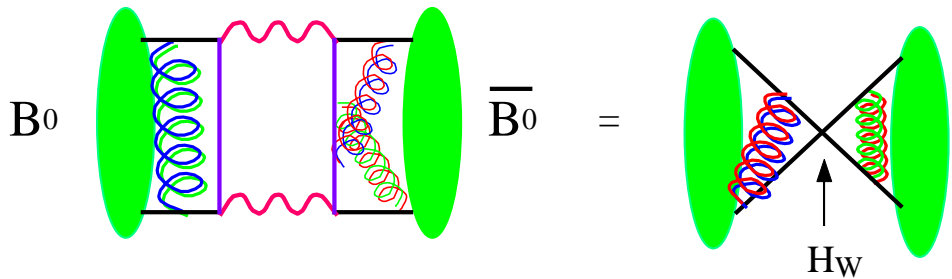}
\caption{\label{fig:bosc}The `box' diagram for the mixing of neutral $B$ mesons 
with their antimeson. On the right is the process that must be calculated 
in lattice QCD.}
\end{figure}

The lattice calculation of $f_B$ is important also for another reason - because 
it is the first stage on the way to calculating the oscillation rate for 
neutral $B$ mesons that will yield the CKM elements in the bottom row of 
the matrix, $V_{td}$ and $V_{ts}$. Neutral $B$ mesons have a meson and 
antimeson that can change into each other via a process dominated 
by the diagram shown in 
Figure~\ref{fig:bosc} in which 
the $\overline{b}$ antiquark in the $B^0$ emits a $W$ and changes to 
$\overline{t}$ antiquark which annihilates with the $d$ quark so that 
another $W$ is created - the 2 $W$ bosons then convert back to 
a $b$ and a $\overline{d}$ which is the antimeson, $\overline{B}^0$. 
This means that the weak 
interactions mix the two states and, as these states propagate in 
time their identity fluctuates, rather like the way energy sloshes 
back and forth between two coupled pendulums. This oscillation can 
be picked up experimentally by studying the decay products from 
a $B^0$, $\overline{B}^0$ pair that are produced in a correlated 
way (e.g. from $e^+e^-$ annihilation at a $B$-factory). This year 
CDF have measured the oscillation rate for the neutral $B_s$ meson 
for the first time~\cite{cdfbs}. This is 
much harder to see, being more rapid, than the rate for the 
$B_d$ meson that has been known for some while. 

\begin{figure}[h]
\begin{minipage}{18pc}
\includegraphics[width=18pc]{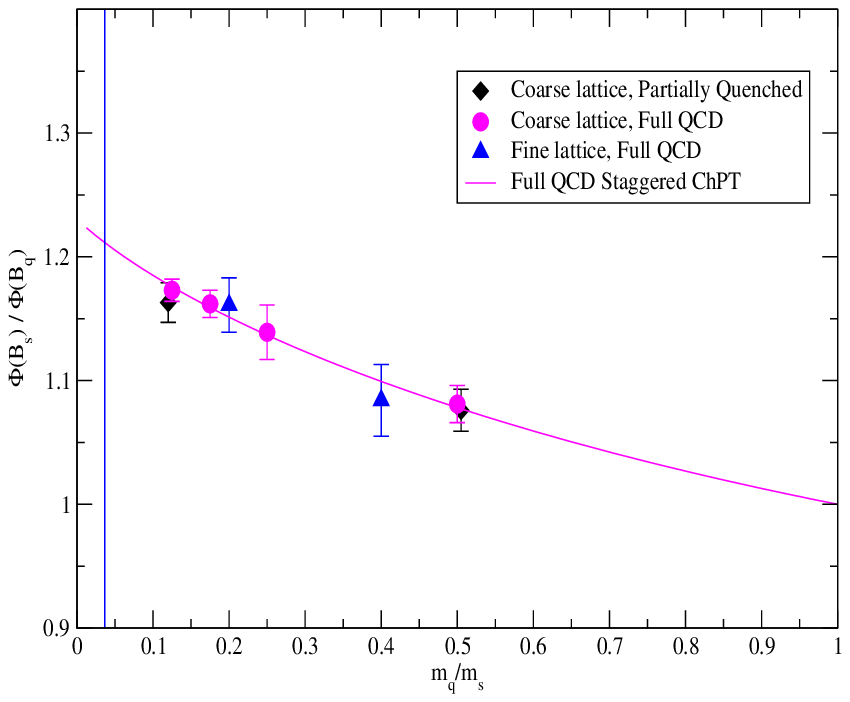}
\caption{\label{fig:fB}The ratio $f_{B_s}\sqrt{M_{B_s}}/f_B\sqrt{M_B}$ as a 
function of the light ($u/d$) quark mass, $m_q$ in units of $m_s$ from 
lattice QCD calculations including the effect of sea quarks~\cite{fB}. The 
calculations have been done at light enough values of $m_q$ that it is 
now possible to extrapolate to the physical answer with only a small error. }
\end{minipage}\hspace{2pc}%
\begin{minipage}{18pc}
\includegraphics[width=18pc]{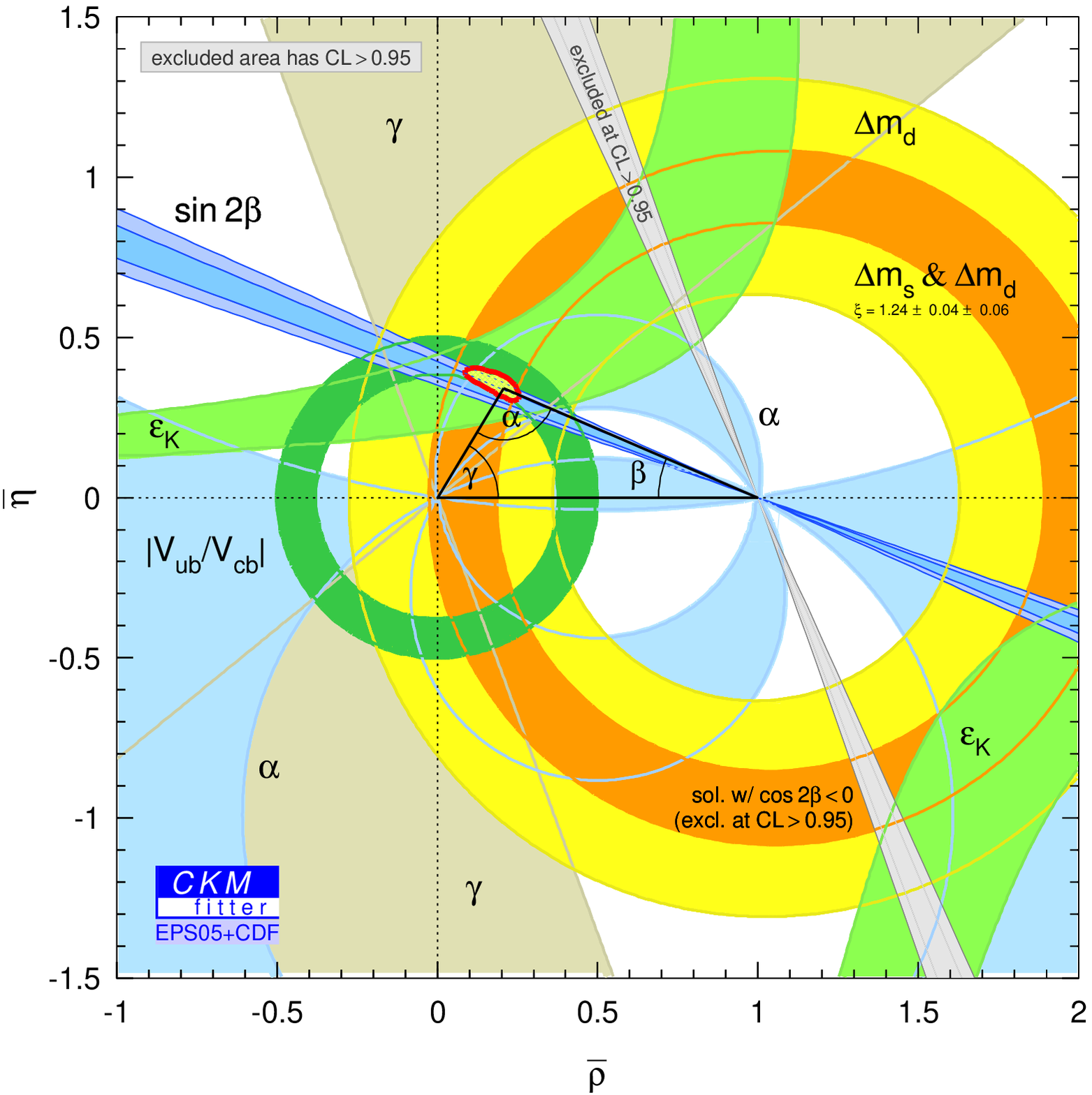}
\caption{\label{fig:ckm}The current status of the search for the 
vertex of the `unitarity triangle'
that expresses the unitarity of the CKM matrix~\cite{ckmfitter}.}
\end{minipage} 
\end{figure}

From Figure~\ref{fig:bosc} it is clear that the oscillation rate will 
depend on the combination $V_{td}V^*_{tb}$ for the $B^0$ meson and 
$V_{ts}V^*_{tb}$ for the $B_s$ meson. The ratio will then yield 
$V_{ts}/V_{td}$ if the rates for the basic process are known, including 
the QCD effects. These can be calculated in lattice QCD from the righthand 
diagram of Figure~\ref{fig:bosc} where the propagation of the heavy $W$ 
and $t$ quarks has been shrunk to an effective point interaction. 
This calculation is ongoing but the initial phase of it (which we believe may 
actually give a good approximation to the answer) is to calculate the 
ratio of the decay constants of the $B_s$ and $B^0$ (although neither can 
actually decay leptonically since they are neutral particles). 

Lattice QCD calculations have now given an accurate result for 
$f_{B_s}/f_{B^0}$ of 1.20(3)~\cite{fB}. Several sources of lattice QCD systematic 
error cancel in this ratio, making it a good quantity to study. 
The result is a significant improvement over previous calculations 
because the effects of sea quarks are included and results are 
available at very light values of $m_{u/d}$. Figure~\ref{fig:fB}
shows the lattice results for several different values of 
$m_{u/d}$. It can be seen that there is significant dependence 
on this mass but that the extrapolation to get 
to the physical answer from the lightest points is now small. 

The results for elements of the CKM matrix can be graphically 
represented as a search for the vertex of a triangle in the 
so-called `$\rho, \eta$' plane where $\rho$ and $\eta$ are a 
parameterisation of the elements of the CKM matrix. The current 
situation from the CKMfitter group~\cite{ckmfitter} is shown in Figure~\ref{fig:ckm}.
The limits provided by the circles of $V_{ub}/V_{cb}$ and 
$\Delta m_s/\Delta m_d$ (the $B_s$ and $B^0$ oscillation rates) 
are very important to pinning down the vertex labelled by angle 
$\alpha$. The accuracy of these constraints is limited by 
the accuracy of the accompanying lattice QCD calculations at 
present~\cite{okamoto, wingate, mackenzie}. The work currently underway, and described here, 
as well as that on semileptonic decay rates~\cite{bsemi, dsemi},
will be key to reducing the size of the blob in the Figure 
to the level where it provides a very serious test of the 
unitarity of the matrix and the self-consistency of the 
Standard Model. 

\section{Conclusions}

Lattice QCD is entering a very exciting period as the first accurate 
calculations that include the effects of sea quarks reach maturity. 
The list of hadron masses and mass differences that agree with experiment 
at the few percent level continues 
to grow, the parameters of QCD have been determined with precision
and a prediction for the mass of the $B_c$ successfully made.
Work continues on the decay rates needed for the accurate determination, 
when combined with experiment, of 
the CKM matrix. There is plenty more to do in lattice QCD and doubtless there 
will be in Beyond the Standard Model 
physics when we find it! 

\ack I am grateful to the conference organisers for the opportunity to present this 
work to such an interesting audience and to my collaborators in the HPQCD and UKQCD 
collaborations.

\smallskip

\end{document}